\documentclass[twocolumn,english]{IEEEtran}
\usepackage[T1]{fontenc}
\usepackage[latin9]{inputenc}
\usepackage{babel}
\usepackage{amsthm}
\usepackage{amsmath}
\usepackage{amssymb}
\usepackage{float}
\usepackage{graphicx}
\usepackage{epsfig}
\usepackage{balance}
\usepackage{multirow,epstopdf}
\usepackage[unicode=true,
 bookmarks=true,bookmarksnumbered=true,bookmarksopen=true,bookmarksopenlevel=1,
 breaklinks=false,pdfborder={0 0 0},backref=false,colorlinks=false]
 {hyperref}
\hypersetup{pdftitle={On Heterogeneous Neighbor Discovery in Wireless Sensor Networks},
 pdfauthor={Lin Chen et al.},
 pdfpagelayout=OneColumn,pdfnewwindow=true,pdfstartview=XYZ,plainpages=false}

\makeatletter

\providecommand{\tabularnewline}{\\}

\theoremstyle{plain}
\newtheorem{thm}{\protect\theoremname}
\theoremstyle{definition}
\newtheorem{defn}[thm]{\protect\definitionname}
\theoremstyle{plain}
\newtheorem{lem}[thm]{\protect\lemmaname}

\@ifundefined{definecolor}{\usepackage{color}}{}
\@ifundefined{definecolor}{\usepackage{color}}{}

\providecommand{\definitionname}{Definition}

\providecommand{\lemmaname}{Lemma}
\providecommand{\theoremname}{Theorem}

\providecommand{\definitionname}{Definition}

\providecommand{\lemmaname}{Lemma}
\providecommand{\theoremname}{Theorem}

\graphicspath{{pix/}}
\usepackage{caption}
\usepackage{subcaption}

\providecommand{\definitionname}{Definition}
\providecommand{\lemmaname}{Lemma}
\providecommand{\theoremname}{Theorem}

\providecommand{\definitionname}{Definition}
\providecommand{\lemmaname}{Lemma}
\providecommand{\theoremname}{Theorem}

\makeatother

\providecommand{\definitionname}{Definition}
\providecommand{\lemmaname}{Lemma}
\providecommand{\theoremname}{Theorem}

\begin{document}

\title{\global\long\def\lcm{\textrm{lcm}}
 \global\long\def\gcd{\textrm{gcd}}
On Heterogeneous Neighbor Discovery \\ in Wireless Sensor Networks}

\author{Lin Chen$^{1,2}$, Ruolin Fan$^3$, Kaigui Bian$^1$, Lin Chen$^4$, Mario Gerla$^3$, Tao Wang$^1$, and Xiaoming Li$^1$\\
    $^1$Peking University, \{abratchen, bkg, wangtao, lxm\}@pku.edu.cn\\
    $^2$Yale University, lin.chen@yale.edu\\
    $^3$University of California, Los Angeles, \{ruolinfan, gerla\}@cs.ucla.edu\\
    $^4$University of Paris-Sud, chen@lri.fr\\
	}

\maketitle
\begin{abstract}

Neighbor discovery plays a crucial role in the formation of wireless sensor networks and mobile networks where the power of sensors (or mobile devices) is constrained. Due to the difficulty of clock synchronization, many asynchronous
protocols based on wake-up scheduling have been developed over the years in order to enable timely neighbor discovery between neighboring sensors while saving energy.
However, existing protocols are not fine-grained enough to support all \emph{heterogeneous} battery duty cycles, which can lead
to a more rapid deterioration of long-term battery health for those without support. Existing research can
be broadly divided into two categories according to their neighbor-discovery techniques---the quorum based
protocols and the co-primality based protocols.
In this paper, we propose two neighbor discovery protocols, called \emph{Hedis} and \emph{Todis}, that optimize the duty cycle granularity of quorum and co-primality based protocols respectively, by enabling the finest-grained control of heterogeneous duty cycles. We compare the two optimal protocols via analytical and simulation results, which show that although the optimal
co-primality based protocol (Todis) is simpler in its design, the optimal quorum based protocol (Hedis) has a better
performance since it has a lower relative error rate and smaller discovery delay, while still allowing the sensor
nodes to wake up at a more infrequent rate.
\end{abstract}
\begin{IEEEkeywords}
Neighbor discovery, heterogeneous duty cycles.
\end{IEEEkeywords}

\section{Introduction}

As human technology continues to advance at an unprecedented rate, there are more mobile wireless devices
in operation than ever before. Many have taken advantage of the ubiquity of these devices to create mobile social network
applications that use mobile sensing as an important feature \cite{Miluzzo:2008}\cite{Pietilainen:2009}. These applications
rely on their devices' capability to opportunistically form decentralized networks as needed.
For this to happen, it is important for these devices to be able to
discover one another to establish a communication link. In order to save
energy, each of the devices alternates between active and sleeping states by keeping its
radio ``ON'' for only some of the time~\cite{feeney2001investigating}.
This is challenging to achieve because two nodes can communicate
only when both of their radios are ``ON'' at the same time; and with clock drifts, having set times for all the
nodes to wake up at the same time is not trivial. Since clock synchronization is difficult in a
distributed system, neighbor discovery must be done asynchronously. Over the years, the
asynchronous neighbor discovery problem has been widely
studied~\cite{searchlight}\cite{disco}\cite{jiang2005quorum}\cite{uconnect}\cite{Zhang:2012:AGO:2426656.2426674}\cite{6257980}\cite{Zheng:2003:AWA:778415.778420},
and existing research mainly focused on satisfying the following three design requirements:
\begin{enumerate}
\item{Guarantee neighbor discovery within a reasonable time frame;}
\item{Minimize the number of time slots for which the node is awake to save energy;}
\item{Match the nodes' awake-sleep schedules with their heterogeneous battery duty cycles as closely as possible to prolong overall battery lifetime\footnote{A duty cycle is the percentage of one period in which a sensor/radio is active.}.}
\end{enumerate}

Most existing solutions to this problem use patterned wake-up
schedules to satisfy the first two requirements. We classify these solutions into two broad categories: (1) \emph{quorum} based protocols that arrange the radio's time slots into a matrix and pick wake-up times according to quorums in the matrix; and (2) \emph{co-primality} based protocols that use numerical analysis to choose numbered time slots as the radio's wake-up times.

In a quorum based protocol, a node populates time slots into a matrix, where the elements in the matrix represent time slots the node takes to run a period of the wake-up schedule \cite{tseng2003power}. The specific arrangements of rows and columns depends upon the protocol scheme, which typically assign slots as ``active'' or ``sleeping'', such that it will ensure these chosen active time slots in the matrix of one node will overlap with those active ones of a neighboring node. Especially, when nodes have the same duty cycles, two nodes choosing active times from a row and a column respectively in the
matrix will be ensured to achieve neighbor discovery regardless of clock drifts.

A co-primality based protocol directly takes advantage of properties of the Chinese Remainder Theorem (CRT)~\cite{nathanson} to ensure that any two nodes would both be active in the same time slot~\cite{disco}. Under these protocols, nodes wake up at time slots in multiples of chosen numbers (a.k.a. protocol parameters) that are co-prime to one another. Such a neighbor discovery protocol fails when nodes choose the same number that would compromise
the co-primality. Thus, every node is allowed to choose several numbers and wake up at multiples of all of those chosen numbers, which guarantees that nodes discover one another within a bounded time/delay.

Up to now, all of the protocols incepted, be it quorum based or co-primality based, fail to meet the third design requirement,
as their requirements for duty cycles are too specific. As a quorum based protocol, Searchlight~\cite{searchlight} requires
that the duty cycles be in the form $\frac{2}{n^{i}}$, where $n$ is a fixed integer (it only supports duty cycles of $1,\frac{1}{2},
\frac{1}{4},\frac{1}{8},\frac{1}{16},\ldots$ if $n=2$). Therefore, it greatly restricts the choices of supported duty cycles due to
the requirement of power-multiples of $2/n$. For a co-primality based protocol like Disco~\cite{disco}, it restricts duty cycles
to be in the form $\frac{1}{p_{1}}+\frac{1}{p_{2}}$, where $p_1$ and $p_2$ are prime numbers. Such stringent requirements
on duty cycles force devices to operate at duty cycles that they are not designed to operate at, thus shortening their
longevity.

In this paper, we present two optimal neighbor discovery protocols, called \emph{Hedis} (\textit{HEterogeneous DIScovery} as a quorum based protocol) and \emph{Todis} (\emph{Triple-Odd based DIScovery} as a co-primality based protocol), that guarantee asynchronous neighbor discovery in a heterogeneous environment, meaning that each node could operate at a different duty cycle.
Specifically, they optimize the duty cycle granularity in their respective protocol categories to support duty cycles in the form of $\frac{2}{n}$ and $\frac{3}{n}$ respectively, and $n$ is an integer that help achieve \emph{almost all }duty cycles smaller than one.
We analytically compare these two protocols with existing state-of-the-art protocols to confirm their
optimality in the support of duty cycles, and also compare them against each other as a comparison between the two general
categories of neighbor discovery protocols (quorum vs. co-primality based protocols). Our results show that while the discovery latencies are
similar for both protocols, Hedis as an optimal quorum based protocol matches actual duty cycles much more closely than Todis as a co-prime based protocol.

The rest of this paper is organized as follows. We formally define the problem as well as any necessary terms in
section~\ref{sec:problem}, and give a taxonomy of current research efforts in this area in section~\ref{sec:background}. In sections~\ref{sec:hedis} and \ref{sec:todis}, we present our optimizations for the quorum based and co-primality based protocols respectively, and we evaluate them with simulations in section \ref{sec:Performance-Evaluation}. Finally, we conclude with section \ref{sec:conclusion}.

\section{Problem Formulation}
\label{sec:problem}
Here we define the terms and variables used to formally describe the neighbor discovery problem and its
solution, as well as state any assumptions used in devising our protocols.

\textbf{Wake-up schedule}. We consider a time-slotted wireless sensor
network where each node is energy-constrained. The nodes follow a
\emph{neighbor discovery wake-up schedule} that defines the time pattern
of when they need to wake up (or sleep), so that they can discover their
respective neighbors in an energy-efficient manner.
\begin{defn}
The neighbor discovery \emph{schedule} (or simply schedule) of a node $a$ is
a sequence $\mathbf{s}_{a}\triangleq\{s_{a}^{t}\}_{0\leq t<T_{a}}$
of period $T_{a}$ and
\[
s_{a}^{t}=\begin{cases}
0 & \text{\ensuremath{a}~sleeps in slot \ensuremath{t}}\\
1 & \text{\ensuremath{a}~wakes up in slot \ensuremath{t}}
\end{cases}.
\]

\textbf{Clock drift}. We do not assume clock synchronization among nodes, therefore
any two given nodes may have random clock drifts. We use the cyclic
rotation of a neighbor discovery schedule to describe this phenomenon. For example,
a clock drift by $k$ slots of node a's schedule $s_a$ is
\[
rotate(\mathbf{s}_{a},k)=\{r_{a}^{t}\}
_{0\leq t<T_{a}},
\]
where $r_{a}^{t}=s_{a}^{(t+k)\bmod T_{a}}$.
\end{defn}

\begin{defn}
The \emph{duty cycle }$\delta_{a}$ of node~$a$ is the percentage of time
slots in one period of the wake-up schedule where node~$a$
is active (node~$a$ wakes up), defined as
\[
\delta_{a}=\frac{|\{0\leq t<T_{a}:s_{a}^{t}=1\}|}{T_{a}}.
\]

\end{defn}

\begin{figure}
\centering
\begin{subfigure}[b]{0.9\linewidth}
	\includegraphics[width=\textwidth]{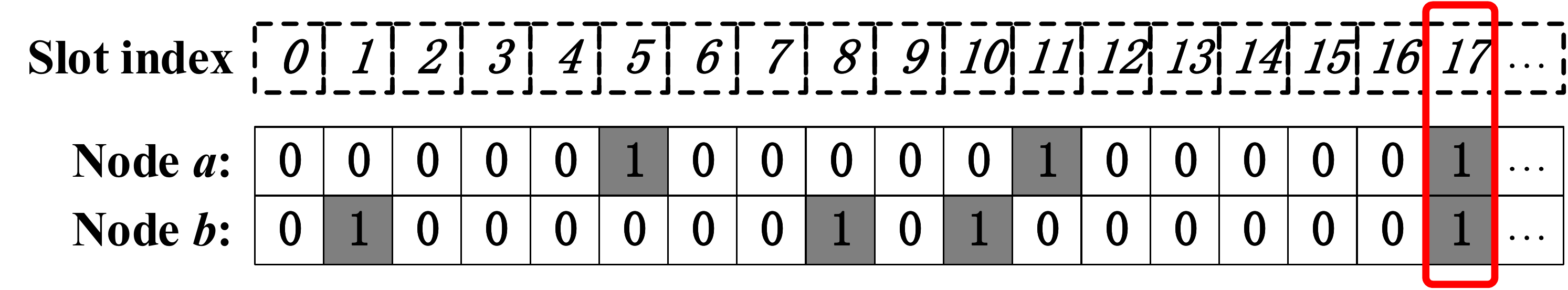} 
	\caption {Without clock drift}
	\label{discovered}
\end{subfigure}
\begin{subfigure}[b]{0.9\linewidth}
	\includegraphics[width=\textwidth]{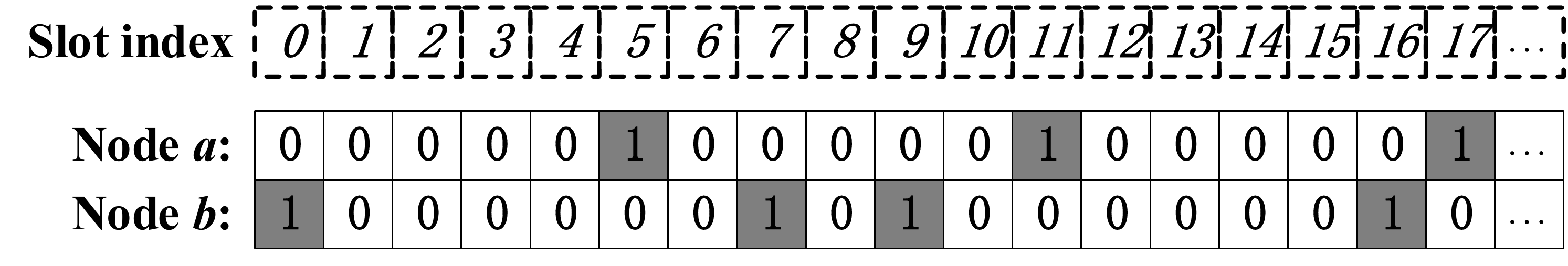} 
	\caption {Node $b$ drifts by 1 time slot to the left}
	\label{missed}
\end{subfigure}
\caption{An example of neighbor discovery: Two neighbor discovery schedules are $\mathbf{s}_{a}=\{0,0,0,0,0,1\}$
and $\mathbf{s}_{b}=\{0,1,0,0,0,0,0,0,1\}$.  Without clock drift (\ref{discovered}), the two nodes can discover each other every 18 time slots since $\lcm(T_{a},T_{b})=18$. With clock drift (\ref{missed}), neighbor discovery fails.}
\label{fig:eg1}
\end{figure}

\textbf{Neighbor discovery}. Suppose two nodes $a$ and $b$ have
schedules $\mathbf{s}_{a}$ and $\mathbf{s}_{b}$ of periods $T_{a}$
and $T_{b}$, respectively. If $\exists t\in[0,\lcm(T_{a},T_{b}))$
such that $\mathbf{s}_{a}^{t}=\mathbf{s}_{b}^{t}=1$ where $\lcm(T_{a},T_{b})$
is the least common multiple of $T_{a}$ and $T_{b}$, we say that:
\begin{itemize}
  \item Nodes $a$ and $b$ can discover each other in slot $t$.
  \item Slot $t$ is called the \emph{discovery slot} between $a$ and $b$.
\end{itemize}
Figure~\ref{fig:eg1} shows an example of two nodes with neighbor discovery
schedules $\mathbf{s}_{a}=\{0,0,0,0,0,1\}$ and $\mathbf{s}_{b}=\{0,1,0,0,0,0,0,0,1\}$,
that have period lengths of $T_{a}=6$ and $T_{b}=9$ respectively.
Node $a$ is active on 1 slot within each period (6 slots) while node
$b$ is active on 2 slots within each period (9 slots). Thus the duty
cycles of $a$ and $b$ are $d_{a}=\frac{1}{6}\approx16.7\%$ and
$d_{b}=\frac{2}{9}\approx22.2\%$. In Figure~\ref{discovered}, We see that for every period of 18 slots ($\lcm(T_{a},T_{b})=18$),
nodes~$a$ and $b$ discover each other in slot 17. However, as illustrated in Figure~\ref{missed},
when a one-slot clock drift occurs in node~$b$, we have $rotate(\mathbf{s}_{b},1)=\{1,0,0,0,0,0,0,1,0\}$
and these two nodes can no longer discover each other.

\section{A Taxonomy of Neighbor Discovery Protocols}
\label{sec:background}
In this section, we introduce a taxonomy of deterministic asynchronous neighbor
discovery protocols. Through examining existing solutions to the neighbor discovery problem,
we divide these protocols into two broad categories.

\subsection{Why Deterministic Protocols}
Many solutions have been proposed to solve the neighbor discovery problem.
One of the earliest such solutions are the birthday protocols \cite{mcglynn2001birthday},
which take upon a \emph{probabilistic} approach to neighbor discovery. These protocols rely on
the \textit{birthday paradox}, which states that with as few as 23 people, the probability that
two people have the same birthday exceeds $\frac{1}{2}$. As a non-deterministic protocol
based upon probability, birthday protocols are heterogeneous and supports every duty
cycle with the finest granularity. Following this, many more similar probabilistic protocols were also developed
\cite{1498535}\cite{Vasudevan:2009:NDW:1614320.1614341}\cite{Zeng:2011:NDW:2107502.2107506}\cite{4524313}.
 However, due to their probabilistic nature, these protocols fail to provide
a guaranteed upper bound for neighbor discovery latency, which means that there is a
chance for two neighbors to never discover each other.

To combat this insufficiency, \emph{deterministic} protocols with worst case bounds for neighbor
discovery were developed. The earlier deterministic
protocols such as \cite{jiang2005quorum}\cite{tseng2003power}, and \cite{lai2010heterogenous} all
use the quorum concept. However, while these protocols are effective in guaranteeing neighbor
discovery, they are generally lacking in duty cycle support. For example, \cite{tseng2003power} and
\cite{jiang2005quorum} are homogeneous, meaning that they require all the nodes to have the
same duty cycle. As a result, the co-primality based approach
was developed with Disco \cite{disco} and U-Connect \cite{uconnect}, although U-Connect is in
some ways a hybrid approach using elements from both the quorum and co-primality paradigms.

\subsection{Quorum vs. Co-primality Based Protocols}
The deterministic protocols for neighbor discovery can be largely classified into two major categories,
\textit{quorum} based protocols and \textit{co-primality} based protocols, though they both work because
of the CRT.

\subsubsection{Quorum Based Protocols}
Quorum based protocols take advantage of geometry in a 2-dimensional array.

\textbf{Bounded discovery delay}. In the most original
protocols like \cite{tseng2003power}, time is arranged into an $m\times m$ matrix.
Every node then chooses a row and a column for
which to wake up. This ensures that regardless of any clock drifts, any two nodes would be able to wake
up at the same time slot at least twice every $m^2$ time slots, thus guaranteeing an upper bound for
neighbor discovery. However, this method only works if every node happens to use the same duty cycle.
Lai et al.~\cite{lai2010heterogenous} improve upon this method by constructing \textit{cyclic quorum
system} and \textit{grid quorum system} pairs, which allow for two different duty cycles to coexist and
still ensure bounded neighbor discovery.

\textbf{Example protocols}. The current latest development in quorum based protocols is Searchlight \cite{searchlight}, which is able
to support multiple duty cycles in the network. Searchlight essentially divides the duty cycle period into
a $\frac{t}{2}\times t$ matrix, and uses a combination of \textit{anchor} and \textit{probing} slots to
generate wake up patterns. At the beginning of every $t$ time slots is an anchor slot, and a probing slot
occurs at random slots between the anchor slots.
With this technique, Searchlight~\cite{searchlight} shows that it is able to
allow neighbor discovery among nodes with many different duty cycles.

\subsubsection{Co-primality Based Protocols}
A co-primality based neighbor discovery protocol is one in which
\begin{itemize}
\item Each node, say, node $a$, chooses a set of integers (not necessarily
distinct) $N_{a}=\{n_{1}^{a},n_{2}^{a},n_{3}^{a},\ldots,n_{|N_{a}|}^{a}\}$.
\item For two distinct nodes $a$ and $b$, $N_{a}$ and $N_{b}$ must satisfy
the following \emph{co-prime pair property}. \end{itemize}
\begin{defn}
For two distinct nodes $a$ and $b$ under a co-primality based neighbor
discovery protocol, there exists an integer in $N_{a}$ that is co-prime
to an integer in $N_{b}$---i.e., $\exists n_{i_{0}}^{a}\in N_{a}$
and $n_{j_{0}}^{b}\in N_{b}$ such that $n_{i_{0}}^{a}$ and $n_{j_{0}}^{a}$
are co-prime.
\end{defn}
Node $a$'s schedule $\mathbf{s}_{a}\triangleq\{s_{a}^{t}\}_{0\leq t<T_{a}}$
under this co-primality based protocol is
\[
s_{a}^{t}=\begin{cases}
1 & t\,\mbox{is divisible by some \ensuremath{n_{i}^{a}\in N_{a}}}\\
0 & \text{otherwise}
\end{cases}.
\]
The period length is $T_{a}=\lcm(n_{1}^{a},n_{2}^{a},\ldots,n_{|N_{a}|}^{a})$
and its duty cycle $\delta_{a}$ is
\[
\delta_{a}=\sum_{1\leq i_{1}\leq|N_{a}|}\frac{1}{n_{i_{1}}^{a}}-\sum_{1\leq i_{1}<i_{2}\leq|N_{a}|}\frac{1}{\lcm(n_{i_{1}}^{a},n_{i_{2}}^{a})}
\]

\[
\cdots+(-1)^{|N_{a}|+1}\frac{1}{\lcm(n_{1}^{a},n_{2}^{a},n_{3}^{a},\cdots,n_{|N_{a}|}^{a})}.
\]

\textbf{Bounded discovery delay}. By the Chinese Remainder Theorem (CRT) \cite{nathanson}, we can obtain the following theorem.
\begin{thm}
\label{thm:main}A co-primality based neighbor discovery protocol
can guarantee discovery for any two nodes for any amount clock drift
if the associated integer sets of the nodes in this network satisfy
the co-prime pair property. And the worst-case discovery delay is
bounded by the product of the two smallest co-prime numbers, one
from each set, i.e.:
\[
\min_{\gcd(n_{i}^{a},n_{j}^{b})=1,1\leq i\leq N_{a},1\leq j\leq N_{b}}\{n_{i}^{a}\cdot n_{j}^{b}\}.
\]

\end{thm}
Suppose the clock of node $a$ is $d$ time slots ahead of that of
node $b$, i.e., node $b$'s $t^{th}$ time slot is the $(t+d)^{th}$
time slot of node $a$, where $d$ is the clock drift,
the following congruence system w.r.t. $t$ applies:
\begin{equation}
\begin{cases}
t\equiv0 & \pmod{p_{i}}\\
t+d\equiv0 & \pmod{p_{j}}
\end{cases}.\label{eq:congruence}
\end{equation}
 If $t$ is a solution to Eq.~(\ref{eq:congruence}), then node $a$
will discover node $b$ in node $a$'s $t$-th time slot (i.e., node
$b$'s $(t+d)$-th time slot). By CRT, since
$p_{i}$ and $p_{j}$ are co-prime, there exists a solution $t\equiv t_{0}\pmod{p_{i}p_{j}}$.

\textbf{Example protocols}.
Disco \cite{disco}, as such a co-primality based protocol,
ensures co-primality by only using prime numbers as possible parameters.
In Disco, each node chooses two distinct
primes to create its wake-up schedule. For example, node $a$ chooses
two distinct primes $p_{1}$ and $p_{2}$ and node $b$ chooses $p_{3}$
and $p_{4}$. Node $a$ is active (wakes up) in the $t$-th time slot
iff $t$ is divisible by either $p_{1}$ or $p_{2}$ while node $b$
is active in the $t$-th time slot iff $t$ is divisible by either
$p_{3}$ or $p_{4}$. Therefore, Disco can guarantee neighbor discovery for any two nodes
for any amount of clock drift with a bounded discovery delay of
\[
\min_{\gcd(p_{i},p_{j})=1,i=1,2,j=3,4}\{p_{i}\cdot p_{j}\}.
\]
Again, this delay is the product of the two smallest co-prime numbers following from the
CRT.

U-Connect \cite{uconnect} is a combination of Disco and the basic quorum based protocol in that it restricts the dimensions of the square quorum matrix to be a prime number. In this way, if the duty cycles of the nodes happen to be the same, neighbors would discover one another via the quorum method. On the other hand, if they are different, the numbers chosen would be co-prime to each other and thus enabling neighbor discovery by Theorem~\ref{thm:main}.

More comprehensive surveys on neighbor discovery can be found in \cite{survey_anastasi} and \cite{survey-Galluzzi}.

\section{Hedis: Optimizing the Quorum based Protocols}
\label{sec:hedis}

Hedis is an asynchronous periodic slot-based neighbor discovery protocol where each node
picks its anchor and probing slots according to the elements of a quorum that is carefully selected in an $(n-1)$ by $n$ matrix.

\subsection{Design of the Hedis Schedule}
For a node~$a$ that has a
desired duty cycle $\delta$, the period of its schedule under Hedis,
$\mathbf{s}_{a}=\{s_{a}^{t}\}_{0\leq t<n(n-1)}$, consists of $n(n-1)$
time slots, where the integer $n$ is chosen such that $\frac{2}{n}$ comes
as close to $\delta$ as possible (and we call $n$ the \emph{parameter}
of this node). Under Hedis, its schedule is
\[
s_{a}^{t}=\begin{cases}
1 & t=ni,(n+1)i+1\text{ (}i=0,1,2,\ldots,n-2\text{)}\\
0 & \text{otherwise}
\end{cases},
\]
 where $ni$ ($i=0,1,2,\ldots,n-2$) denotes the index of an \emph{anchor}
slot and $(n+1)i+1$ denotes the index of a \emph{probing} slot.

\begin{figure}
\centering
\begin{subfigure}[b]{0.8in}
    \includegraphics[width=\textwidth]{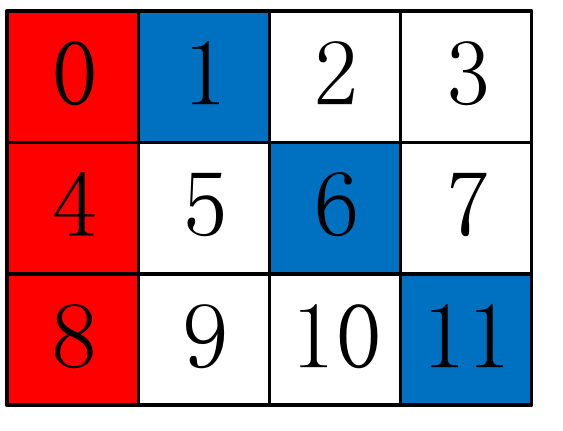}
    \caption{n=4}
    \label{fig:n4}
 \end{subfigure}
 \hspace{0.1in}
 \begin{subfigure}[b]{1in}
    \includegraphics[width=\textwidth]{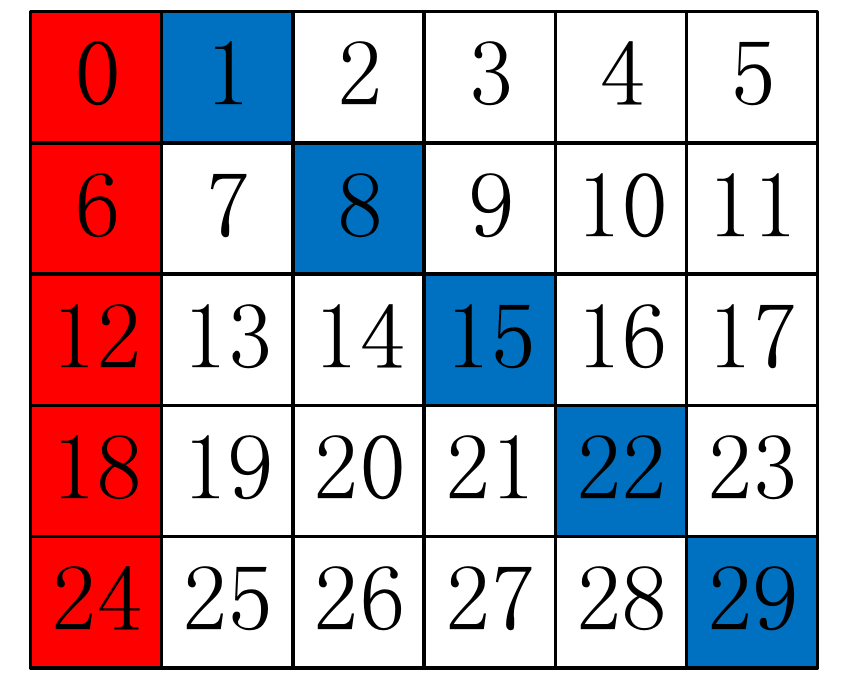}
    \caption{n=6}
    \label{fig:n6}
\end{subfigure}
\caption{Two example Hedis schedules.}
\label{fig:hedis}
\end{figure}

\begin{figure*}
 \centering \includegraphics[width=7in]{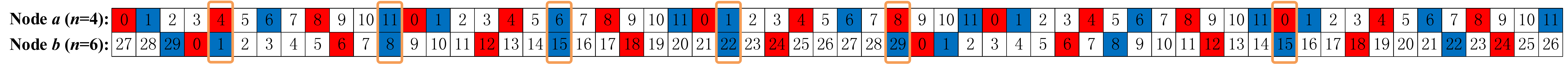} \caption{Node~$a$'s schedule is 3-slot ahead of node~$b$'s schedule. The
overlapped colored slots between their schedules represent the successful
neighbor discovery.}
\label{fig:m=00003D4_n=00003D5}
\end{figure*}

Figure~\ref{fig:hedis} shows two example Hedis schedules when $n=4,6$,
and the two schedules consist of of $n(n-1)=12,30$ time slots, respectively.
Each grid in the figure represents a time slot, and the integer inside
a grid denotes its slot index, e.g., the grid with $0$ inside denotes
the 0th time slot in the schedule (note that a schedule starts from
the 0th time slot). The red and blue slots represent the anchor and
probing slots, during which the node wakes up. When $n=4$, the duty cycle is $2/4=50\%$. The full schedules are depicted in
Figure~\ref{fig:m=00003D4_n=00003D5}, where the two nodes with different
duty cycles can achieve successful neighbor discovery (overlap of
colored slots between schedules of nodes~$a$ and $b$) for many
times in every period. Next, we will show that Hedis can guarantee
neighbor discovery for any two nodes of same-parity parameters (both
odd or both even) with heterogeneous duty cycles for any amount of
clock drift.

\subsection{Bounded Discovery Delay under Hedis}
\begin{lem}
\label{lem:nathan}Let $m$ and $n$ be positive integers. For any
integers $a$ and $b$, there exists an integer $x$ such that
\begin{equation}
x\equiv a\pmod{m}\label{eq:xeqa}
\end{equation}
 and
\begin{equation}
x\equiv b\pmod{n}\label{eq:xeqb}
\end{equation}
 if and only if
\[
a\equiv b\pmod{\text{gcd}(m,n)}.
\]
 If $x$ is a solution of congruences (\ref{eq:xeqa}) and (\ref{eq:xeqb}),
then the integer $y$ is also a solution if and only if
\[
x\equiv y\pmod{\text{lcm}(m,n)}.
\]

\end{lem}
The proof can be found in \cite{nathanson}, on page 61, Theorem 2.9.
By Lemma~\ref{lem:nathan}, we further establish the following theorem.
\begin{thm}
\label{thm:Hedis-can-guarantee} Hedis guarantees neighbor discovery
within bounded latency for any two nodes with the same-parity parameters
$n$ and $m$, given any amount of clock drift between their schedules.
The average discovery latency is $O(nm)$. \end{thm}
\begin{IEEEproof}
Nodes $a$ and $b$ are two arbitrarily given nodes, whose parameters
are $n$ and $m$, respectively. The periods of the Hedis schedules
of nodes $a$ and $b$ are $T_{a}=n(n-1)$ and $T_{b}=m(m-1)$, respectively.
We use $d$ to denote the clock drift.

Without loss of generality, we study the following system of congruences
with respect to $t$:
\begin{equation}
\begin{cases}
t\equiv ni+d,(n+1)i+1+d\pmod{n(n-1)}\\
t\equiv mj,(m+1)j+1\pmod{m(m-1)},
\end{cases}\label{eq:system_of_congruences}
\end{equation}
 where $i\in[0,n-2]$, $j\in[0,m-2]$.
\[
t\equiv ni+d,(n+1)i+1+d\pmod{n(n-1)}
\]
($i\in[0,n-2]$) is true iff $\exists i\in[0,n-2]$ such that it is
true, and the same meaning for
\[t\equiv mj,(m+1)j+1\pmod{m(m-1)}
\]
($j\in[0,m-2]$).

There are a number of $nm$ pairs of simultaneous congruences, which we divide into 4 groups:
anchor-anchor, anchor-probing, probing-anchor and probing-probing groups. E.g., the anchor-probing group
denotes the case where an anchor slot of node $a$ overlaps a probing slot of node $b$. Note that if we find a
solution that meets the requirements of any one of these congruences, we obtain a solution to
Eq.~(\ref{eq:system_of_congruences}).

\textbf{Group~1: anchor-anchor}. Consider the following system of
congruences
\[
\begin{cases}
t\equiv ni+d\pmod{n(n-1)} & i\in[0,n-2]\\
t\equiv mj\pmod{m(m-1)} & j\in[0,m-2]
\end{cases},
\]
 which is equivalent to
\begin{equation}
\begin{cases}
t\equiv d\pmod{n}\\
t\equiv0\pmod{m}
\end{cases}.\label{eq:case1}
\end{equation}
 By Lemma \ref{lem:nathan}, Eq.~(\ref{eq:case1}) has a solution
if and only if
\[\gcd(n,m)\mid d.
\]

\textbf{Group~2: anchor-probing}. Consider the following system of
congruences
\begin{equation}
\begin{cases}
t\equiv ni+d\pmod{n(n-1)} & i\in[0,n-2]\\
t\equiv(m+1)j+1\pmod{m(m-1)} & j\in[0,m-2]
\end{cases},\label{eq:case2-2}
\end{equation}
 which is equivalent to
\begin{equation}
\begin{cases}
t\equiv d\pmod{n}\\
t\equiv(m+1)j+1\pmod{m(m-1)} & j\in[0,m-2]
\end{cases}.\label{eq:case2}
\end{equation}
 By Lemma \ref{lem:nathan}, Eq.~(\ref{eq:case2}) has a solution
if and only if $\gcd(n,m(m-1))\mid(m+1)j+1-d$ for some integer $j\in[0,m-2]$,
i.e., the congruence with respect to $j$
\begin{equation}
(m+1)j\equiv d-1\pmod{\gcd(n,m(m-1))}\label{eq:case2-3}
\end{equation}
 has a solution.

\textbf{Group~3: probing-anchor}. Consider the following system of
congruences
\begin{equation}
\begin{cases}
t\equiv(n+1)i+1+d\pmod{n(n-1)} & i\in[0,n-2]\\
t\equiv mj\pmod{m(m-1)} & j\in[0,m-2]
\end{cases},\label{eq:case3-1}
\end{equation}
 which is equivalent to
\begin{equation}
\begin{cases}
t\equiv(n+1)i+1+d\pmod{n(n-1)} & i\in[0,n-2]\\
t\equiv0\pmod{m}
\end{cases}.\label{eq:case3}
\end{equation}
 By Lemma \ref{lem:nathan}, Eq.~\ref{eq:case3} has a solution if
and only if
\[
\gcd(m,n(n-1))\mid(n+1)i+1+d
\]
for some integer $i\in[0,n-2]$,
i.e., the congruence with respect to $i$
\[
(n+1)i\equiv-d-1\pmod{\gcd(m,n(n-1)}
\]
 has a solution.

\textbf{Group~4: probing-probing}. Consider the following system
of congruences
\begin{equation}
\begin{cases}
t\equiv(n+1)i+1+d\pmod{n(n-1)} & i\in[0,n-2]\\
t\equiv(m+1)j+1\pmod{m(m-1)} & j\in[0,m-2]
\end{cases}.\label{eq:case4}
\end{equation}
 By Lemma \ref{lem:nathan}, Eq.~\ref{eq:case4} has a solution if
and only if
\[
\gcd(n(n-1),m(m-1))\mid(n+1)i-(m+1)j+d
\]
for some integer
$i\in[0,n-2]$ and $j\in[0,m-2]$.

Now we begin to prove this theorem by cases.

\textbf{Case~1:} If $m>n$, the congruence system of anchor-probing (Group 2)
is true. \textit{Proof:} If $m>n$, we have $m-1\geq n\geq\gcd(n,m(m-1))$. And note
that $\gcd(m+1,\gcd(n,m(m-1)))=\gcd(m+1,n,m(m-1))=\gcd(m+1,2,n)=1$.
This is because $m$ and $n$ are both odd or are both even. So one
of $m+1$ and $n$ are odd, and we have $\gcd(m+1,2,n)=1$. Therefore
$(m+1)j$ ($j\in[0,m-2]$) runs over all congruence classes modulo
$\gcd(n,m(m-1))$. Then Eq.~(\ref{eq:case2-3}) has at least $\lfloor(m-1)/\gcd(n,m(m-1))\rfloor$
solutions and on average $(m-1)/\gcd(n,m(m-1))$ solutions. Hence
Eq.~(\ref{eq:case2}) has at least $\lfloor(m-1)/\gcd(n,m(m-1))\rfloor$
solutions and on average $(m-1)/\gcd(n,m(m-1))$ solutions modulo
$\lcm(n,m(m-1))$. Therefore, the average discovery latency is $\frac{\lcm(n,m(m-1))}{(m-1)/\gcd(n,m(m-1))}=nm$.

\textbf{Case~2:} If $n>m$, the congruence system of probing-anchor (Group 3)
is true. \textit{Proof:} If $n>m$, similarly to case 1, we have Eq.~(\ref{eq:case3})
has at least $\lfloor(n-1)/\gcd(m,n(n-1))\rfloor$ solutions and on
average $(n-1)/\gcd(m,n(n-1))$. Hence Eq.~(\ref{eq:case3-1}) has
at least $\lfloor(n-1)/\gcd(m,n(n-1))\rfloor$ solutions and on average
$(n-1)/\gcd(m,n(n-1))$ modulo $\lcm(m,n(n-1))$. Therefore, the average
discovery latency is $nm$.

\textbf{Case~3}. If $n=m$, we consider the result of $d\bmod n$. If $d\equiv0\pmod{n}$,
then $\gcd(n,m)=n\mid d$, and thus the anchor-anchor case (Group 1) is true
and the average discovery latency is $O(nm)$. Now we concentrate
on the case where $d\not\equiv0\pmod{n}$. Since $n=m$, Eq.~(\ref{eq:case2-3})
becomes
\[
(n+1)j\equiv d-1\pmod{n}.
\]
 Since $(n+1)j=nj+j\equiv j\pmod{n}$, this is equivalent to
\[
d\equiv j+1\pmod{n}.
\]
 For $j\in[0,n-2]$, $j+1$ runs over$[1,n-1]$. Because $d\not\equiv0\pmod{n}$,
there exists a $j\in[0,n-2]$ that satisfies Eq.~(\ref{eq:case2-3}),
and therefore the anchor-probing case (Group~2) is true. Similarly, the probing-anchor
case (Group 3) is also true. And it is easy to check that the average discovery
latency is $O(n^{2})$, i.e., $O(nm)$.
\end{IEEEproof}

\section{Todis: Optimizing Co-primality based Protocols}
\label{sec:todis}

Now we optimize the asynchronous co-primality based protocols, and propose Todis that exploits properties of consecutive odd integers for achieving co-primality.

As a co-primality based protocol, Todis creates wake-up schedules for the nodes
based on multiples of numbers that are co-prime to each other. This ensures that
any two given nodes would be able to wake up at the same time by the co-prime
pair property as illustrated in Section \ref{sec:background}, thus succeeding
in neighbor discovery. Recall that Disco~\cite{disco} guarantees this by simply using prime numbers as parameters, which limits the variety of parameters to choose from.

For two nodes $a$ and $b$, we need to construct two sets of integers, $N_a$ and $N_b$, that must satisfy the co-prime pair property. In our quest to find co-prime pairs, we observe that for two numbers to be co-prime, at least one of them must be odd. Thus, we explore the possibility of achieving co-primality using odd integers. We observe that given two odd integers $a$ and $b$, if they are not co-prime, often times either ``$a+2$ and $b$'', or ``$a$ and $b+2$'' is a co-prime pair. For example, if 15 and 21 are not co-prime, we are able to find that either ``17 and 21'', or ``15 and 23'' is a a co-prime pair. Following this logic, we design our Todis protocol using sets of consecutive integers.

\subsection{Design of the Todis Schedule}

\subsubsection{Trying sets of two consecutive integers}
First, we tried using two consecutive
odd integers in $N_{a}$ for each node~$a$, i.e., $N_{a}=\{n,n+2\}$
where $n\geq1$ and $n$ is odd. Unfortunately, for given nodes~$a$
and $b$, there are instances where the sets $N_{a}$ and $N_{b}$ do not satisfy
the co-prime pair property for very small numbers (i.e., less than 100).
For example, when $N_{a}=\{33,35\}$ and
$N_{b}=\{75,77\}$ and $\forall n_{i}^{a}\in N_{a},n_{j}^{b}\in N_{b}$,
we have $\gcd(n_{i}^{a},n_{j}^{b})>1$.


\subsubsection{Using sets of three consecutive integers in Todis}

In Todis, we use three consecutive odd integers $n-2$, $n$ and $n+2$ ($n\geq3$) for constructing a wake-up schedule.

The co-prime pair property requires that at least one of the three consecutive odd integers
that node $a$ chooses (i.e., $n-2$, $n$ and $n+2$) is co-prime
w.r.t. one of the three integers that node $b$ chooses (i.e., $m-2$,
$m$ and $m+2$).

\textbf{Bounded discovery delay in practical networks}. Generally, the triples consisting of three consecutive odd integers can also fail to satisfy the required co-prime pair property, as seen in
counterexamples shown by the CRT. However,
the two smallest sequences of odd integers in these counterexamples are
$N_{a}=\{1600023,1600025,1600027\}$
and $N_{b}=\{2046915,2046917,2046919\}$. Such integers are too
large to be chosen for creating a ``practical'' duty cycle anyway.
For example, an $n$ value larger than $1600023$ would
imply a duty cycle smaller than $\delta_{a}$ of $0.00000187496$.
In practical applications, however, duty cycles are
much greater than $0.00000187496$. Therefore, any chosen sets $N_{a}$
and $N_{b}$ based on duty cycles would satisfy the co-prime pair property.
By Theorem \ref{thm:main}, Todis guarantees neighbor discovery with
a delay bounded by
\[
\min_{\gcd(n+i,m+i)=1,i=-2, 0, 2}\{n\cdot m\}.
\]


A node~$a$ that has a desired duty cycle of $\delta$ may therefore choose an odd integer $n$ such that
\[
\frac{3(n^{2}-n-1)}{n(n^{2}-4)}\approx\frac{3}{n} = \hat{\delta}
\]
 is as close to $\delta$ as possible. We call $n$ the \emph{parameter}
of node $a$.

Under Todis, its wake-up schedule is
\[
s_{a}^{t}=\begin{cases}
1 & t~\mbox{is divisible by either \ensuremath{n-2}, \ensuremath{n}, or \ensuremath{n+2}}\\
0 & \text{otherwise}
\end{cases},
\]
with a period length of $(n-2)n(n+2)$ and a duty cycle of
\[
\frac{1}{n-2}+\frac{1}{n}+\frac{1}{n+2}-\frac{1}{(n-2)n}-\frac{1}{n(n+2)}-\frac{1}{(n-2)(n+2)}
\]

\[
+\frac{1}{(n-2)n(n+2)}=\frac{3(n^{2}-n-1)}{n(n^{2}-4)}.
\]

\begin{figure}
 \centering \includegraphics[width=4cm]{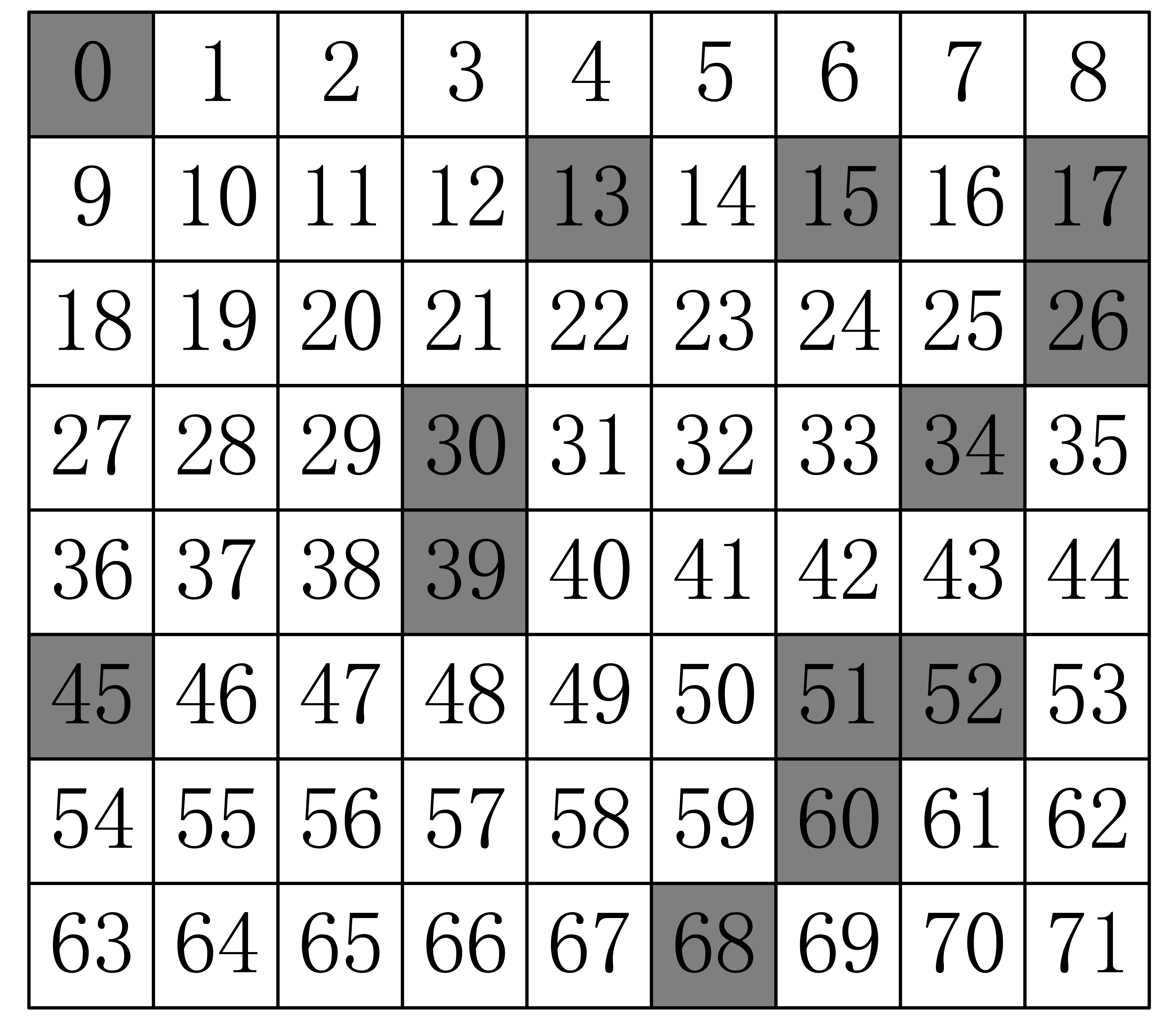}\caption{The first 71 time slots under the Todis schedule when $n=15$ (i.e.,
the node chooses $13$, $15$ and $17$). The node wakes up in time slots
0, 13, 15, 17, 26, 30, 34, 39, 45, 51, 52, 60, 68, $\ldots$.\label{fig:eg_todis}}
\end{figure}

Figure~\ref{fig:eg_todis} shows the first 71 time slots under the Todis
schedule when $n=15$ (i.e., the node chooses $13$, $15$ and $17$).
Each grid in the figure represents a time slot, and the integer inside
a grid denotes its slot index, e.g., the grid with $0$ inside denotes
the 0th time slot in the schedule (note that a schedule starts from
the 0th time slot). The gray slots represent the active slots where
the node wakes up. In this example, the duty cycle is $\frac{3\cdot(5^{2}-5-1)}{5\cdot(5^{2}-4)}\approx18.9\%$.

\subsection{Analysis of Duty Cycle Granularity}
Now we discuss the granularity of Todis in matching any desired duty cycle in practical applications.
Suppose node $a$'s desired duty cycle is $\delta_{a}$, the relative error $\epsilon(\delta_{a})$ between
$\delta_{a}$ and its approximation $\hat{\delta_{a}}$ is defined by
\begin{equation}
  \label{eqn:err}
  \epsilon(\delta_{a})=\frac{\left|\hat{\delta}_{a}-\delta_{a}\right|}{\delta_{a}}.
\end{equation}
We want to mathematically estimate the upper bound of $\epsilon$ given
$\delta_{a}$, which we denote as $\hat{\epsilon}(\delta_{a})$.

In Todis, node $a$ needs to
choose an odd integer $n_{a}$ such that $\frac{3(n_{a}^{2}-n_{a}-1)}{n_{a}(n_{a}^{2}-4)}$
lies closest to $\delta_{a}$, i.e.,
\[
n_{a}=\mathrm{arg}\min_{n\text{ odd}\mbox{\text{ }}}|\frac{3(n^{2}-n-1)}{n(n^{2}-4)}-\delta_{a}|.
\]
Thus, the best approximation of the desired duty cycle $\delta_{a}$
is
\[
\hat{\delta}_{a}=\frac{3(n_{a}^{2}-n_{a}-1)}{n_{a}(n_{a}^{2}-4)}\equiv\min_{n\text{ odd}}|\frac{3(n^{2}-n-1)}{n(n^{2}-4)}-\delta_{a}|.
\]

Let $f(2k-1)$ and $f(2k+1)$ be two consecutive supported duty cycles,
where $f(n)=\frac{3(n^{2}-n-1)}{n(n^{2}-4)}$. Relative error $\epsilon$
reaches a local maximum at $\delta_{a}=\frac{f(2k-1)+f(2k+1)}{2}$.
Thus we obtain a quartic equation with respect to $k$
\begin{equation}
16\delta_{a}k^{4}-24k^{3}+(12-40\delta_{a})k^{2}+36k+9\delta_{a}-9=0.\label{eq:wrt-k}
\end{equation}
By Eq.~(\ref{eq:wrt-k}), we can obtain a solution $k=k(\delta_{a})$ in complex radicals
(the other three solutions are discarded). Then we have
\[
\epsilon(\delta_{a})\leq\hat{\epsilon}(\delta_{a})\triangleq\frac{f(2k(\delta_{a})-1)-\delta_{a}}{\delta_{a}},
\]
 where $\hat{\epsilon}(\delta_{a})$ is also a complex expression
in radicals with respect to $\delta_{a}$.

\begin{figure*}[ht]
    \begin{minipage}[t]{0.3\textwidth}
     \centering
         \includegraphics[width=2.1in]{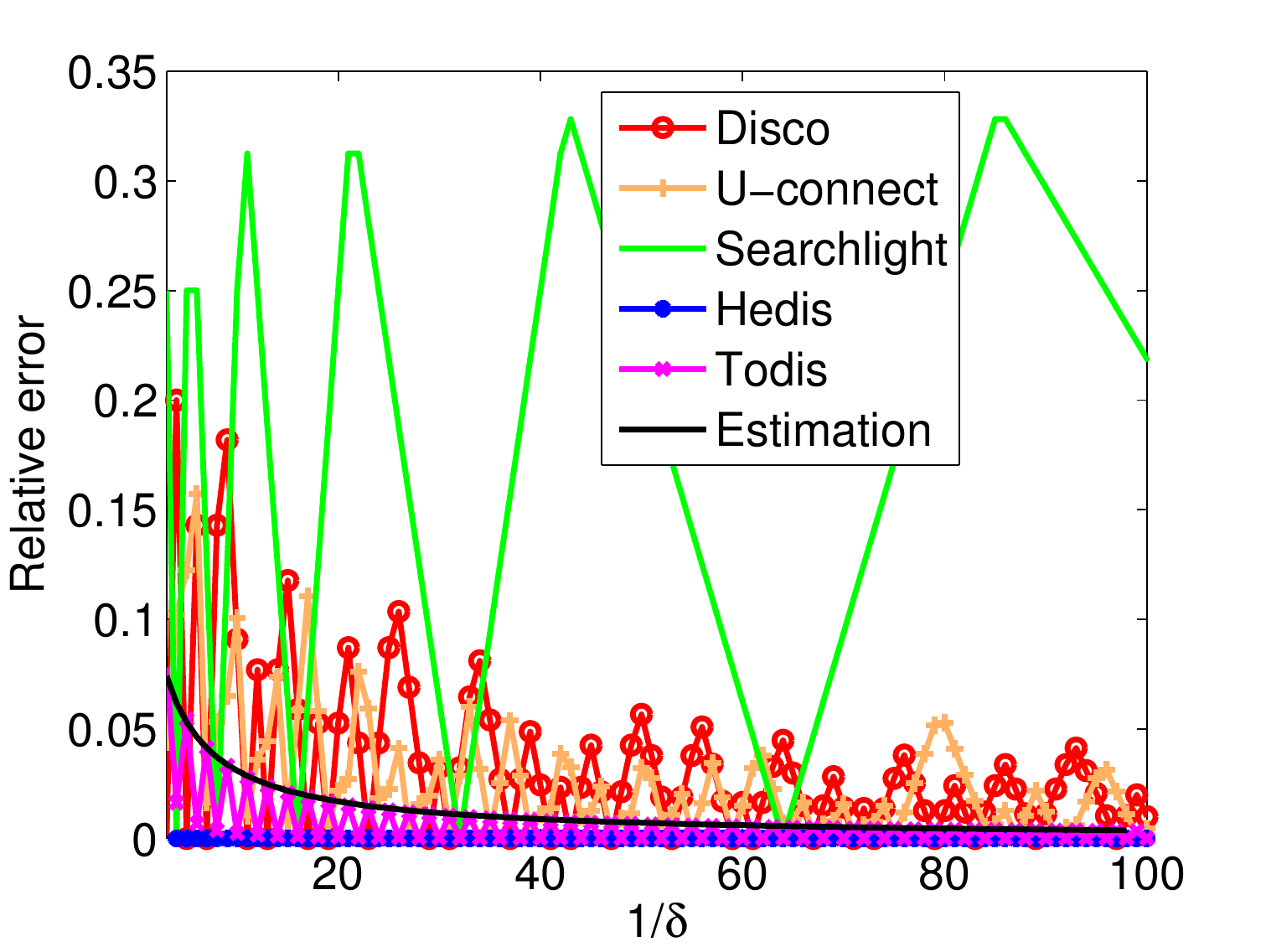} \caption{Relative error vs. small duty cycle $\delta$.
         The ``Estimation" line is the theoretical upper bound estimation of relative error induced by Todis (see
         Section \ref{sec:todis}).}
        \label{fig:small}
    \end{minipage}
\hfill
 \begin{minipage}[t]{0.3\textwidth}
     \centering
        \includegraphics[width=2.1in]{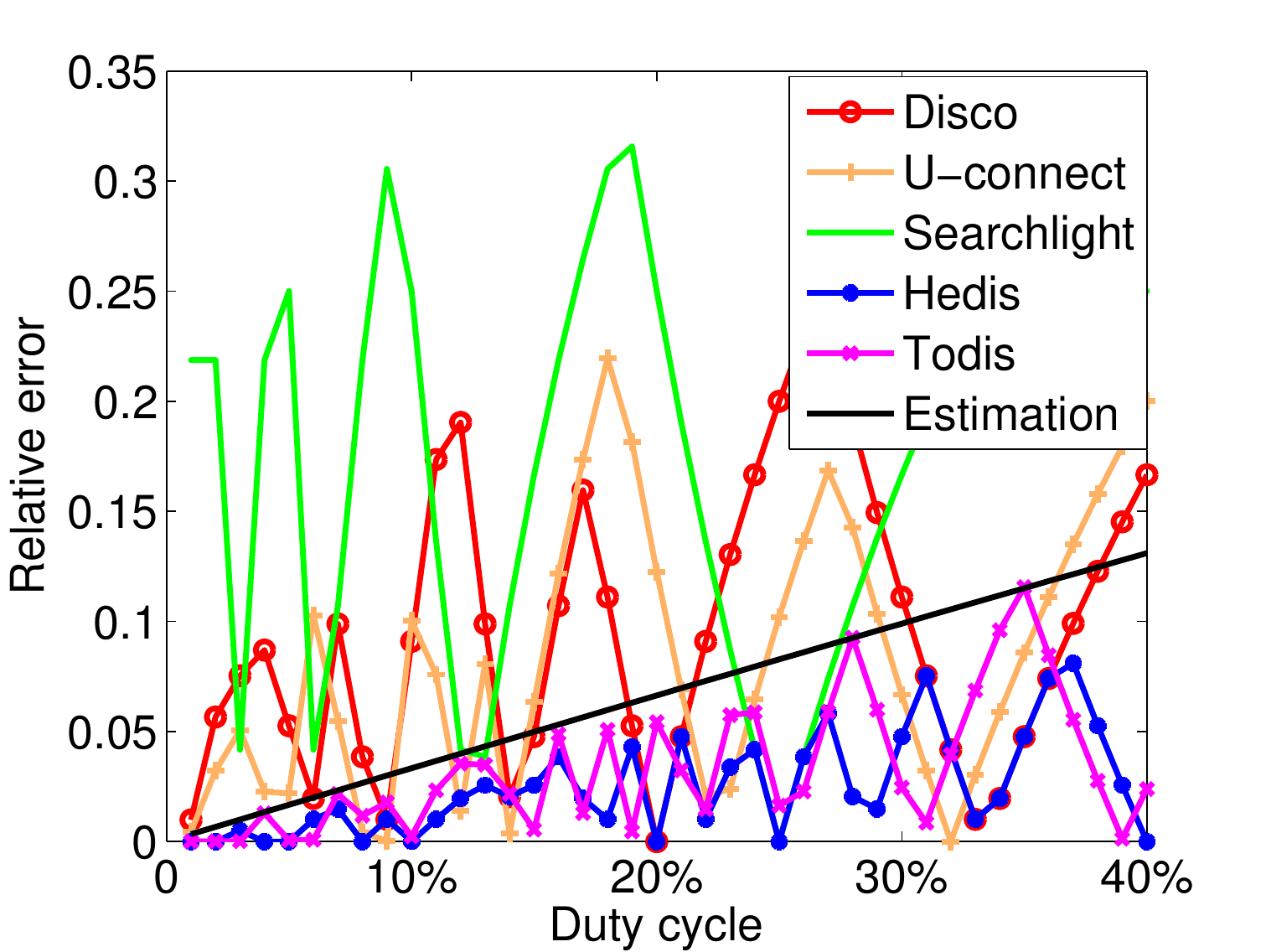}
        \caption{Relative error vs. large duty cycle $\delta$.
         The ``Estimation" line is the theoretical upper bound estimation of relative error induced by Todis (see
         Section \ref{sec:todis}).}
        \label{fig:large}
    \end{minipage}
\hfill
  \begin{minipage}[t]{0.3\textwidth}
     \centering
      \includegraphics[width=2.1in]{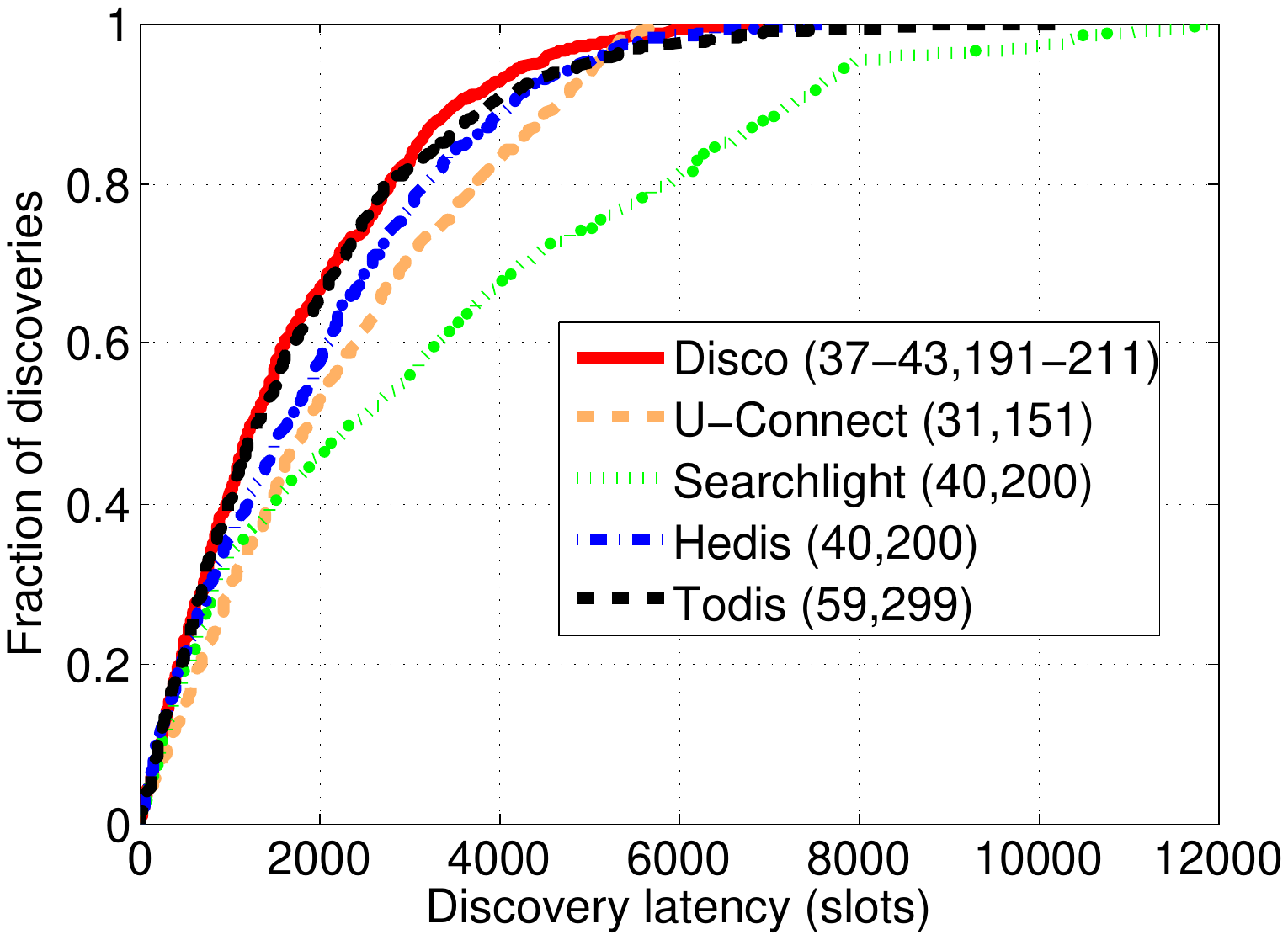}
        \caption{CDF of discovery latency when each pair of nodes operate at duty cycles
        $1\%$ and $5\%$, respectively. Numbers in the parentheses indicate the parameters of each corresponding protocol.}
        \label{fig:CDF5}
    \end{minipage}
\end{figure*}

Note that $\epsilon(\delta_{a})=\hat{\epsilon}(\delta_{a})$
iff $\delta_{a}=\frac{3(n^{2}-n-1)}{n(n^{2}-4)}$ for some odd integer
$n$. We illustrate $\hat{\epsilon}(\delta_{a})$ in Figures \ref{fig:small} and \ref{fig:large}
(see the ``Estimation'' lines), and we can observe that $\hat{\epsilon}(\delta_{a})$
is a very tight upper bound for $\epsilon(\delta_{a})$.

The upper bound function $\hat{\epsilon}(\delta_{a})$ is an increasing
function in $[0,1)$. In practical applications, $\delta_{a}$ is smaller
than $20\%$, and thus $\epsilon$ is upper bounded by $6.71\%$,
which is a very small relative error. Moreover, $\epsilon$ drops below $3.34\%$ when $\delta_{a}\leq10\%$. Asymptotically,
\[
\hat{\epsilon}(\delta_{a})\simeq\frac{2\delta_{a}}{\sqrt{9+4\delta_{a}^{2}}+3}\simeq\frac{1}
{3}\delta_{a}
 \]
linearly approaches 0 as $\delta_{a}$ goes to 0. This property implies that the error decreases with the decline of the desired duty cycle.

\section{Performance Evaluation\label{sec:Performance-Evaluation}}

We compare Hedis and Todis against state-of-the-art neighbor discovery protocols
of both the quorum based and the co-primality based varieties.
These protocols include Disco~\cite{disco} (co-primality based),
Searchlight~\cite{searchlight} (quorum based), and U-Connect~\cite{uconnect} (a combination
of both). We evaluate the performances of
these protocols using two metrics, namely the discovery latency and the duty cycle granularity.
\begin{itemize}
  \item In Disco, each node chooses a pair of primes $p_{1}$
        and $p_{2}$ to support duty cycles of the form $\frac{1}{p_{1}}+\frac{1}{p_{2}}$,
        and the worst-case discovery latency is $\min\{p_{1}p_{3},p_{1}p_{4},p_{2}p_{3},p_{2}p_{4}\}$.
  \item In U-Connect, each node wakes up 1 time slot every
        $p$ time slots and wakes up $\frac{p+1}{2}$ time slots every $p^{2}$
        time slots. Therefore U-Connect supports duty cycles of the form
        $\frac{3p+1}{2p^{2}}$, and has the worst-case discovery latency of
        $p_{1}p_{2}$ if one node uses prime $p_{1}$ while another uses $p_{2}$. The dependence of Disco and U-Connect upon prime numbers greatly restricts their support of choices of duty cycle varieties.
  \item Searchlight requires that a node's parameter $n_{1}$ be a multiple or factor of
        its neighboring node's parameter $n_{2}$ to guarantee
        neighbor discovery. Therefore, in a network that implements Searchlight,
        the number that each node chooses must be a power-multiple
        of the smallest chosen number (i.e., 2, 4, 8, 16, or 3, 9, 27, 81,
        etc.), guaranteeing that any two nodes' numbers are multiples of each
        other. As a result, Searchlight only supports duty cycles of the form
        $\frac{2}{t^{i}}$, where $t$ is an integer (i.e., the aforementioned
        smallest chosen number) and $i=0,1,2,3,\ldots$
\end{itemize}

\begin{table}
\centering \caption{Comparison of Hedis and Todis with existing neighbor discovery protocols.}
\small

\label{tab:Comparisons} %
\begin{tabular}{|c||c|c|c|}
\hline
Protocol  & Parameter  & Average  & Supported \tabularnewline
name  & restriction  & dis. delay  & duty cycles\tabularnewline
\hline
\hline
\multirow {2}{*}{Disco}  & prime $p_{1},p_{2}$  & $O(\min\{p_{1}p_{3},$  & \multirow {2}{*} {$\frac{1}{p_{1}}+\frac{1}{p_{2}}$ }\tabularnewline
 & $p_{3},p_{4}$  & $p_{1}p_{4}, p_{2}p_{3},p_{2}p_{4}\})$  & \tabularnewline
\hline
\multirow {2}{*}{U-Connect}   & prime  & \multirow {2}{*}{ $O(p_{1}p_{2})$ }  & \multirow {2}{*}{$\frac{3p_{1}+1}{2p_{1}^{2}}$ }\tabularnewline
 & $p_{1}$ , $p_{2}$  &  & \tabularnewline
\hline
\multirow {2}{*}{Searchlight}  & power-multiple  & \multirow {2}{*} { $O(t_{1}t_{2})$ } & \multirow {2}{*}{ $\frac{2}{t_{1}^i}$ } \tabularnewline
 & of $t_{1}$ , $t_{2}$  &  & \tabularnewline
\hline
\multirow {2}{*}{Hedis}  & same parity  & \multirow {2}{*} { $O(nm)$ }   & \multirow {2}{*} { $\frac{2}{n}$ }\tabularnewline
 & $n$, $m$  &  & \tabularnewline
\hline
\multirow {2}{*}{Todis}	  &  \multirow {2}{*} { odd $n,m$ } & \multirow {2}{*} { $O(nm)$  }  & $\frac{3(n^{2}-n-1)}{n(n^{2}-4)}$ \tabularnewline
           &            &         & $\approx \frac{3}{n}$ \tabularnewline
\hline
\end{tabular}
\end{table}

Table \ref{tab:Comparisons} gives an overall numerical comparison among these protocols. As the table shows, while the
difference in discovery latency exists among these protocols, all of them perform on the order of the multiple of the principle
parameters in the two participating nodes.
\begin{itemize}
  \item Discovery latencies may be similar among the different protocols, because two nodes may choose similar parameters so as to match the desired duty cycle.
  \item In contrast, the metric of duty cycle granularity presents a different story. While all the parameters used in the protocols all have special restrictions due
to protocol design, it is obvious that those for Hedis and Todis are the least stringent. For example, fewer than
$2\%$ of integers under 1000 are prime, while half of them are odd, giving Todis a much larger pool of numbers
to choose from for its parameters as compared to Disco and U-Connect.
\end{itemize}
We confirm these numerical results using simulations. We measure the
relative errors each of the aforementioned protocols yields at differing duty cycles,
as well as their discovery latencies in node pairs operating at various duty cycles.


\subsection{Duty Cycle Granularity}
The first set of simulations comparatively studies the supported duty cycles.
We study two groups of duty cycles:
\begin{enumerate}
\item Small duty cycles $1\leq1/\delta\leq100$, i.e., $\delta=1,\frac{1}{2},\frac{1}{3},\frac{1}{4},\ldots,\frac{1}{100}$;
\item Equispaced large duty cycles $0\leq\delta\leq1$, i.e., $\delta=0\%,1\%,2\%,3\%,4\%,\ldots,100\%$.
\end{enumerate}
We use the metric called \emph{the relative error} (defined in Eq.~\ref{eqn:err}) to quantify the
capability of supporting each studied duty cycle, which is denoted
as
\[
\epsilon\triangleq|\delta'-\delta|/\delta,
\]
 where $\delta'$ is the closest duty cycle that is supported by each
simulated protocol, w.r.t. $\delta$. Note that a smaller $\epsilon$
implies that the protocol provides more choices for energy conservation
with a finer granularity of duty cycle control.
For Searchlight, we let the smallest duty cycle unit be $1/2$ to
allow the finest duty cycle granularity.

Figure~\ref{fig:small} illustrates the results for small duty cycles, while
Figure~\ref{fig:large} shows those of large duty cycles. These results
provide us the following insights:
\begin{itemize}
\item Searchlight is inferior to the other protocols in supporting various duty
cycles because it requires the duty cycle to be $\frac{2}{t^{i}}$,
where $t$ is a fixed integer and $i=1,2,3,\ldots$. In this simulation,
we use $t=2$ to give Searchlight support for the duty cycles $1,\frac{1}{2},\frac{1}{4},\frac{1}{8},\ldots$.
The relative error increases significantly as the desired duty cycle
deviates away from the supported duty cycles (e.g., in Figure~\ref{fig:large},
it has a peak at $37.5\%=\frac{1/2+1/4}{2}$, and $1/2$ and $1/4$
are supported duty cycles).

\item The schedules in Disco and U-Connect are generated using prime
numbers, which have a denser distribution than power-multiples. Thus,
Disco and U-Connect perform better than Searchlight.

\item Both Hedis and Todis greatly outperform all the other protocols, having very small relative errors. In fact, for small duty cycles, the relative errors from Hedis
is nearly constantly zero (see Figure \ref{fig:small}). On the other hand, although
Todis also performs well, its error rate obviously increases much faster than Hedis
as the duty cycle $\delta$ increases.

\item The theoretical ``Estimation" lines for Todis (see Figures~\ref{fig:small} and  \ref{fig:large}) holds up well in that it follows
the same pattern as Todis' actual error rates. This confirms our prior analysis
in section \ref{sec:todis} of Todis' duty cycle granularity, where we estimated the
upper bound relative error for Todis.

\end{itemize}

%
%
%
%
%

\subsection{Discovery Latency}
\begin{figure*}[ht]
    \begin{minipage}[t]{0.3\textwidth}
     \centering
\includegraphics[width=2.1in]{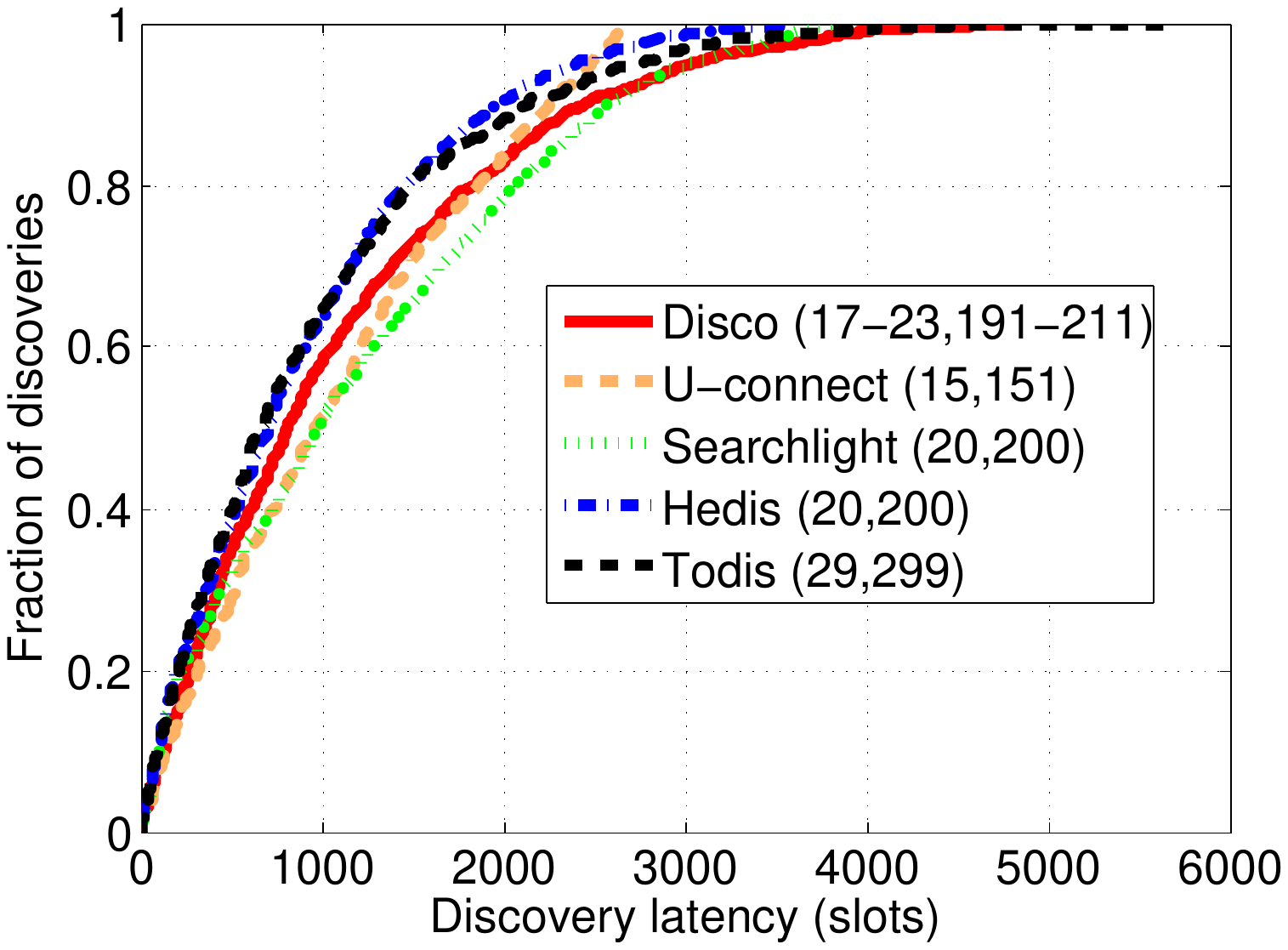}
\caption{CDF of discovery latency when each pair of nodes operate at duty cycles
$1\%$ and $10\%$, respectively. Numbers in the parentheses indicate the parameters of each corresponding protocol}
\label{fig:CDF10}
    \end{minipage}
\hfill
 \begin{minipage}[t]{0.3\textwidth}
     \centering
        \includegraphics[width=2.1in]{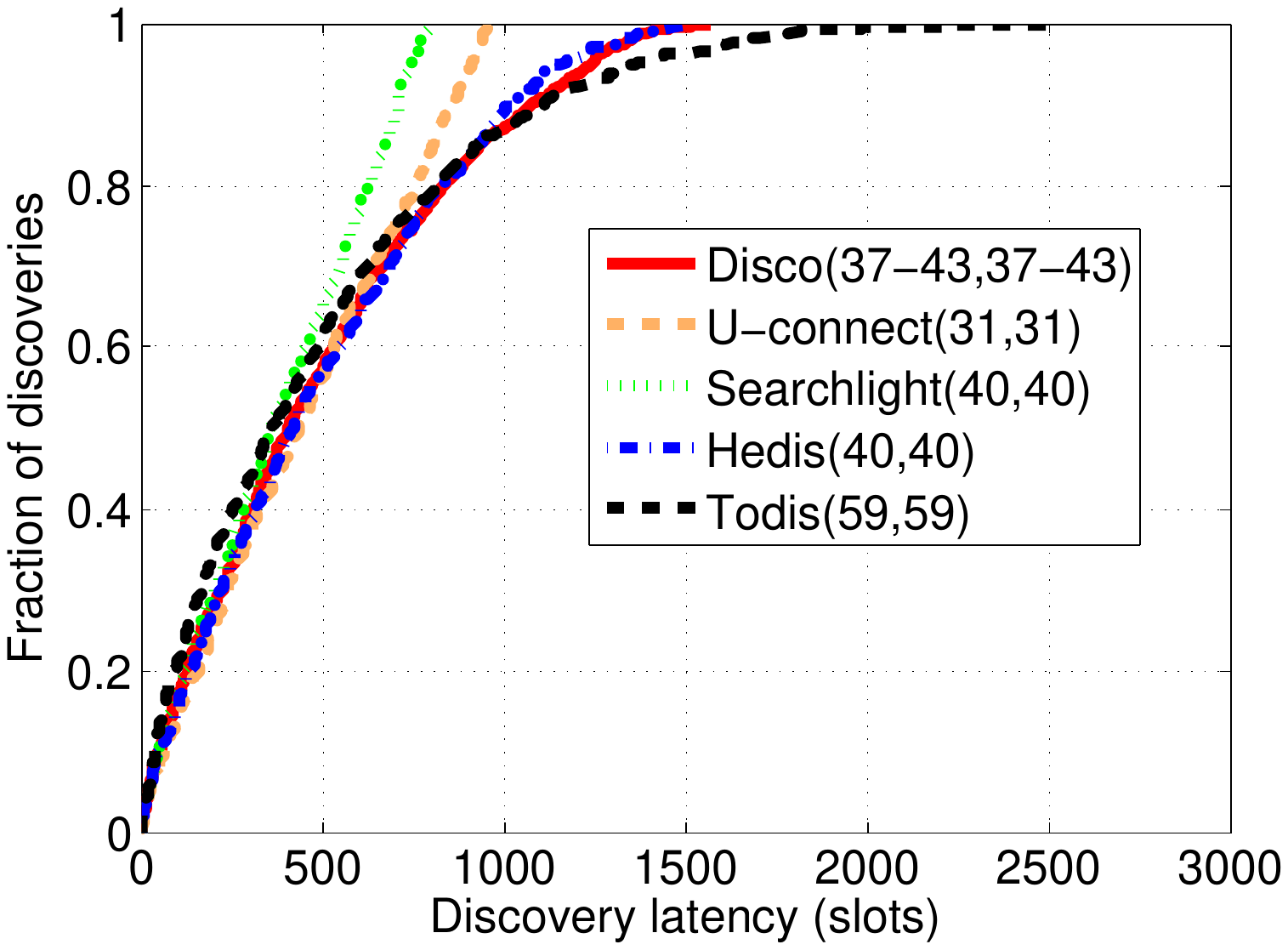}
        \caption{CDF of discovery latency when each pair of nodes operate at the same duty
        cycle of $5\%$. Numbers in the parentheses indicate the parameters of each corresponding protocol.}
        \label{fig:CDF5-5}
    \end{minipage}
\hfill
  \begin{minipage}[t]{0.3\textwidth}
     \centering
      \includegraphics[width=2.1in]{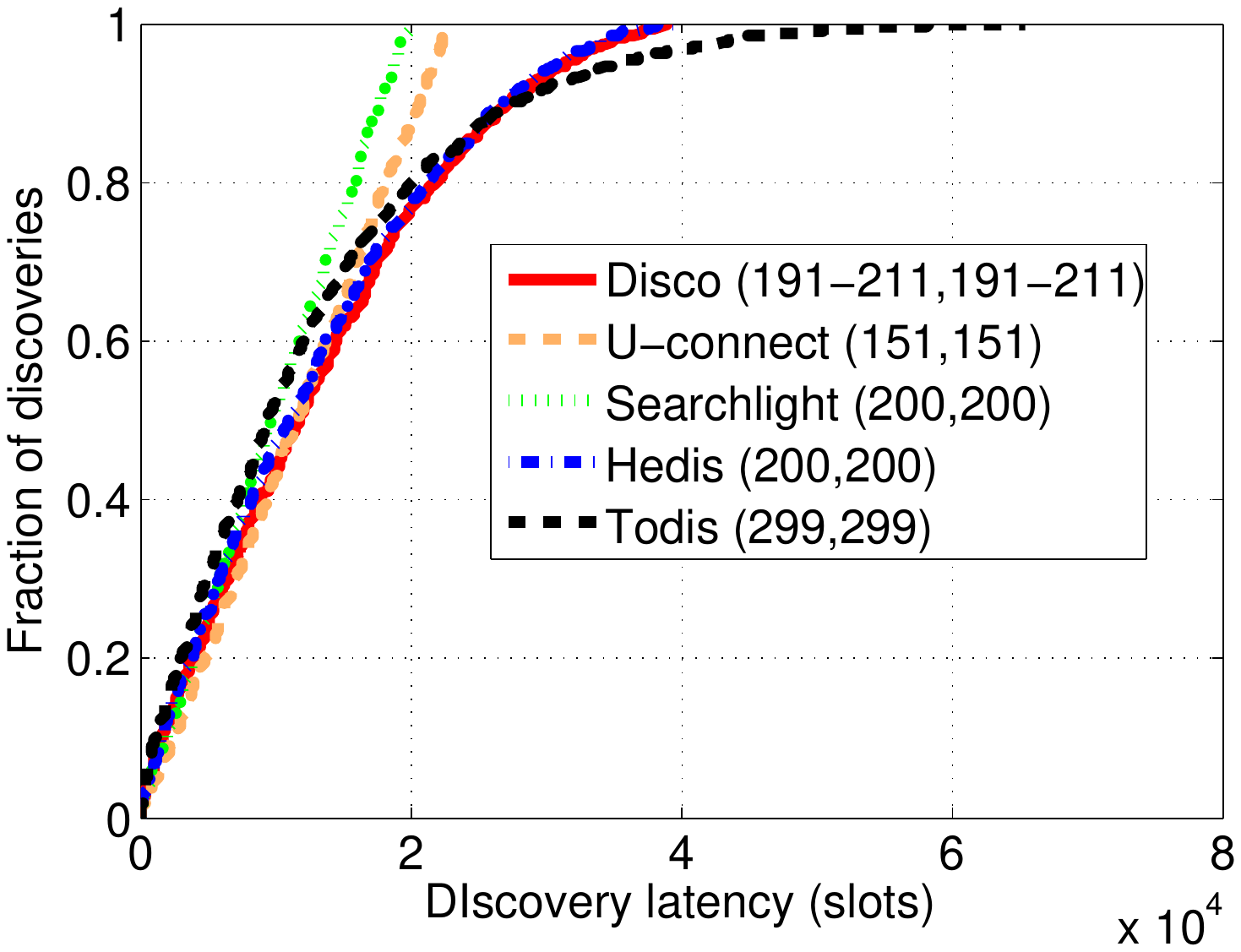}
       \caption{CDF of discovery latency when each pair of nodes operate at the same duty
        cycle of $1\%$. Numbers in the parentheses indicate the parameters of each corresponding protocol.}
        \label{fig:CDF1-1}
    \end{minipage}
\end{figure*}

In this set of simulations, we study the discovery latencies of these protocols. For each simulation, we take 1000 independent
pairs of nodes and assign various duty cycles. In two instances, we compare the protocols' performance in
heterogeneous discovery scenarios. We assign duty cycles of $1\%$ and $5\%$ to each respective node in the node pair in
the first instance (see Figure \ref{fig:CDF5}), and $1\%$ and $10\%$ in the second instance (see Figure
\ref{fig:CDF10}). We also compare the performance of the protocols in two homogeneous discovery scenarios,
with each node in the node pair operating at the same duty cycles of $5\%$ in the first scenario (see Figure \ref{fig:CDF5-5})
and $1\%$ in the second one (see Figure \ref{fig:CDF1-1}).

\textbf{Heterogeneous vs. homogeneous duty cycles}. From these four cumulative distribution function graphs (CDFs), we see that overall, all of the protocols have comparative
discovery latencies, with the odd exception of Searchlight in Figure~\ref{fig:CDF5}. Nonetheless,
it must be noted that all 5 protocols presented were eventually successful in neighbor discovery for $100\%$ of the pairs
tested. These CDFs also show that Hedis is one of the few protocols that consistently perform above average in both
the heterogenous and homogeneous neighbor discovery cases.
For example, Figures~\ref{fig:CDF5-5} and~\ref{fig:CDF1-1} indicate that Searchlight is the clear winner for discovery latency in the homogeneous case, but it does poorly in the heterogeneous cases, as seen in Figures~\ref{fig:CDF5} and \ref{fig:CDF10}.

In addition, we see that for up to $90\%$ of the CDF, Hedis and Todis are both near top
performers, but the protocol with one of the smallest latencies in reaching $100\%$ of the CDF in every case is U-Connect.
We attribute this to the fact that U-Connect uses smaller values as its parameters, thus having a smaller upper bound in the worst case.

Similarly, we attribute Todis' consistent long tail in each CDF scenario to its larger parameters. Therefore, although it can
quickly allow nodes to discover each other in most cases, seen in its quickly reaching $90\%$ in the CDFs, it has the longest latency in the worst-case scenarios.

\textbf{Hedis vs. Todis}. These various simulations show that Hedis and Todis optimize the duty cycle granularity in both the quorum based and the
co-primality based neighbor discovery approaches, with Hedis having a finer granularity than Todis. Additionally, both
protocols perform reasonably well in terms of discovery latency, with Todis having a larger worst case latency bound due to its larger parameters.


%
%

\section{Conclusion}
\label{sec:conclusion}

In this paper, we explored the current two main approaches of designing an asynchronous heterogeneous neighbor discovery
protocol with guaranteed latency upper bounds---the quorum based and the co-primality based approaches. Using
these two approaches we designed the Hedis and Todis neighbor discovery protocols, emphasizing on duty cycle granularity
optimization for both. Hedis, as a quorum based protocol, forms a $(n-1) \times n$ matrix of time slots and uses the anchor-
probing slot method to ensure neighbor discovery. Todis, as a co-primality based protocol, uses sets of three consecutive odd
integers to ensure co-primality and thus ensures neighbor discovery due to CRT. In the design of both
protocols we proved their capability in ensuring acceptable upper bounds in discovery latency. Through analytical
comparisons as well as simulations, we confirmed the optimality of Hedis and Todis in duty cycle granularity among existing
protocols. Hedis is able to support duty cycles in the form of $\frac{2}{n}$, while Todis can support duty cycles roughly in the
form of $\frac{3}{n}$, allowing both protocols to effectively cover any practical duty cycle and thus prolong battery longevity.

We also showed in both our analysis and simulations that Hedis as a quorum based protocol is better than Todis as a
co-primality based protocol in both duty cycle granularity and discovery latency, although differences by the latter metric are
minor. By being able to support duty cycles at such a fine granularity while still guaranteeing an acceptable discovery latency
bound, Hedis truly paves the way for neighbor discovery in wireless sensor networks.

\small
\balance
\bibliographystyle{abbrv}
\bibliography{reference-list}

\end{document}